\documentclass[superscriptaddress, prl, twocolumn]{revtex4}

\usepackage{times}
\usepackage{graphicx}% Include figure files
\usepackage{subfigure}
\usepackage{dcolumn}% Align table columns on decimal point
\usepackage{bm}% bold math
\usepackage{multirow}

\def\ket#1{\mathinner{|{#1}\rangle}}

\begin{document}
\author{Avinash Kolli}
\email{avinash.kolli@materials.ox.ac.uk}
\affiliation{Department of Materials, Oxford University, Oxford OX1 3PH, UK}
\author{Brendon W. Lovett}
\email{brendon.lovett@materials.ox.ac.uk}
\affiliation{Department of Materials, Oxford University, Oxford OX1 3PH, UK}
\author{Simon C. Benjamin}
\affiliation{Department of Materials, Oxford University, Oxford OX1 3PH, UK} 
\author{Thomas M. Stace}
\affiliation{DAMTP, University of Cambridge, Cambridge, CB3 0WA, UK}

\title{All-Optical Measurement Based QIP in Quantum Dots}

\begin{abstract}
Parity measurements on qubits can generate the entanglement resource necessary for scalable quantum computation. Here we describe a method for fast optical 
parity measurements on electron spin qubits within coupled quantum dots. The measurement scheme, which can be realised with existing technology, consists of 
the optical excitation of excitonic states followed by monitored relaxation. Conditional on the observation of a photon, the system is projected into the 
odd/even parity subspaces. Our model incorporates all the primary sources of error, including detector inefficiency, effects of spatial separation and 
non-resonance of the dots, and also unwanted excitations. Through an analytical treatment we establish that the scheme is robust to such effects. Two 
applications are presented: a realisation of a CNOT gate, and a technique for growing large scale graph states. \end{abstract}

\maketitle

Quantum computation (QC) offers the possibility of exponential speed-up over classical computation \cite{nielsen00}. Many of the ideas put forward for 
implementing QC in the solid state involve using electron spins to represent qubits. In order to create controlled entanglement these schemes typically 
envisage some mechanism for switching on and off spin-spin interactions \cite{kane98}. This is an enormous challenge experimentally. Recently, Beenakker \textit{et al.} \cite{beenakker04} have shown 
that it is possible to use parity measurements on pairs of spins rather than interaction switching. Together with single spin rotations, such measurements 
suffice to implement scalable QC. A scheme for exploiting this idea in the solid state with electrostatically defined dots has been advanced by Engel and 
Loss \cite{engel05}; while Barrett and Stace \cite{barrett06} propose a similar scheme for a spin singlet-triplet measurement. These ideas rely on charge 
detection and therefore require electrode structures in the vicinity of the qubits. Here we consider an alternative optical measurement of spin parity that can be implemented in quantum dot (QD) structures.

The optical process involves the excitation of 
odd-parity spin states to higher excitonic states. The readout is achieved by the radiative relaxation of these excited states, and then the observation of a 
photon which projects the system into the odd-parity subspace. Conversely, when no photon is observed the system is projected into the even-parity subspace. Since any additional single qubit gates can also be implemented through optical pulses \cite{lovett06}, this scheme constitutes an all-optical approach to 
measurement based QC in the solid state.

\textit{Model} - Consider two QDs, each of which are \textit{n} doped so that they each contain an excess conduction band electron. The qubit basis 
$|0\rangle$ and $|1\rangle$ is defined by the electron spin states $m_{z}=-1/2$ and $1/2$ respectively. We consider subjecting the structure to a single 
laser radiating both QDs with $\sigma^{+}$ polarised light. By the Pauli blocking effect, the creation of an exciton is possible only for the 
$|m_{z}=1/2\rangle$ state \cite{pazy03}. This exciton-spin (trion) state is denoted as $|X\rangle$.

\begin{figure}[t]
\begin{center}
\includegraphics[width=.5\textwidth ]{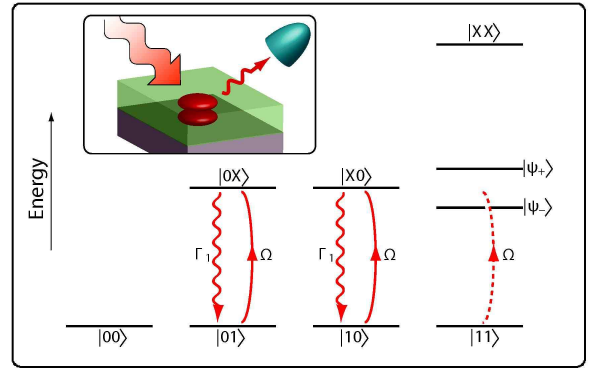}
\vspace{-1cm}
\caption{\label{fig1} 
Inset: schematic of a system suitable for a two-qubit demonstation. The double dot is subjected to an excitation pulse and monitored for photon emission (collection apparatus not shown). Main figure: energy level structure for two resonant dots, showing the operation of the parity measurement. The 
creation of excitons is only possible in the odd-parity subspace. The relaxation process in the odd-parity subspace results in the emission of a photon which is subsequently detected.}
\end{center}
\end{figure}

The Hamiltonian for our two quantum dots driven by a classical laser field is
\begin{eqnarray} \label{eq:H} H(t) &=& \omega_{a} |X\rangle \langle X| \otimes \hat{I} + \omega_{b} \hat{I} \otimes |X\rangle \langle X| + 
V_{XX} |XX\rangle \langle XX| \nonumber\\ && + V_{F}\big(|1X\rangle \langle X1| + H.c.\big)  + \Omega \cos \omega_{l}t \big(|1\rangle \langle X| \otimes 
\hat{I} \nonumber\\ && + \hat{I} \otimes |1\rangle \langle X| + H.c.\big) , \end{eqnarray}  
where H.c. denotes hermitian conjugate and $\omega_{a}$ and $\omega_{b}$ are the exciton creation energies for dot $a$ and dot $b$ respectively. 
$V_{F}$ is the strength of the Foerster interaction, which causes exciton transfer between the dots via virtual photon exchange. $V_{XX}$ 
is the biexcitonic energy shift due to the exciton-exciton dipole interaction, $\Omega$ is the time-dependent laser coupling (assumed to be the same for both 
dots), and $\omega_{l}$ is the laser frequency. We have neglected the energy difference between the $|0\rangle$ and $|1\rangle$ states, as it is negligible 
on the exciton energy scale. The Foerster interaction is non-magnetic and couples only states $|X 1\rangle$ and $|1 X\rangle$. We first consider 
resonant dots such that $\omega_{a}=\omega_{b}=\omega_{0}$ (see Fig.~\ref{fig1}). 

The Hamiltonian (\ref{eq:H}) may be decoupled into three subspaces with no interactions between them: $\mathcal{H}_{0} = \{ |00\rangle \}$, $\mathcal{H}_{1}=\{ |01\rangle ,
|0X\rangle , |10\rangle , |X0\rangle \}$, $\mathcal{H}_{2}=\{ |11\rangle , |X1\rangle, |X1\rangle , |XX\rangle \}$. Let us look at the Hamiltonian for the last of 
these subspaces. We write this in a basis of the eigenstates for $\Omega = 0$, which are $|11\rangle$, $|\psi_{+}\rangle = \frac{1}{\sqrt{2}}(|1X\rangle + 
|X1\rangle)$, $|\psi_{-}\rangle =\frac{1}{\sqrt{2}} (|1X\rangle - |X1\rangle)$ and $|XX\rangle$. The degeneracy of the $|\psi_{-}\rangle$ and $|\psi_{+}\rangle$ 
levels is lifted by the Foerster interaction, resulting in two states each containing a delocalized exciton. In this basis:
\begin{eqnarray} \label{eq:H2} H_{2} &=& (\omega_{0} + V_{F}) |\psi_{+}\rangle \langle\psi_{+}| + (\omega_{0} - V_{F}) |\psi_{-}\rangle \langle \psi_{-}| \nonumber\\ && + 
(2 \omega_{0} + V_{XX})|XX\rangle \langle XX| \nonumber\\ && + \Omega^{'} \cos \omega_{l}t \big(|11\rangle \langle \psi_{+}| + |\psi_{+}\rangle \langle XX| + h.c. \big) .
\end{eqnarray}
Thus, the only dipole allowed transitions in this subspace are between $|11\rangle$ and $|\psi_{+}\rangle$, and between $|\psi_{+}\rangle$ and 
$|XX\rangle$, each with a coupling strength of $\Omega^{'}=\Omega\sqrt{2}$.

We achieve our parity-measurement by applying a $\pi$-pulse tuned to the exciton creation energy $\omega_{0}$, which will populate the $|0X\rangle$ and 
$|X0\rangle$ states fully, while the $|00\rangle$ and $|11\rangle$ states remain as they are owing to the Foerster splitting (see Fig.~\ref{fig1}). Next we allow the system to 
relax: if we measure a photon without determining from which QD it originated, we expect the state of the system to be projected into the spin-parity odd 
subspace while retaining the initial coherence between the $|01\rangle$ and $|10\rangle$ states. If no photon is measured then we expect that the 
system will collapse into the even-parity subspace, again retaining the necessary coherence for the parity-measurement.

%To retain the coherence between the $|01\rangle$ and $|10\rangle$ states when a photon is emitted, we require that it is not possible to distinguish from 
%which of the two QDs the photon originates.

For perfect fidelity of operation we need to ensure that after the initialization procedure there is no population of the 
$|\psi_{+}\rangle$ and $|XX\rangle$ states. Returning to the Hamiltonian for the $\mathcal{H}_2$
subspace (Eq. \ref{eq:H2}), moving into a frame rotating at frequency $\omega_{l} = \omega_{0}$ and making a rotating wave approximation, we may write
\begin{eqnarray} H_{2} &=& -V_{F} |\psi_{-}\rangle \langle \psi_{-}| + V_{F} |\psi_{+}\rangle \langle \psi_{+}| + V_{XX} |XX\rangle \langle XX| \nonumber \\ &&+ 
\Omega^{'}/2 \big(|11\rangle \langle \psi_{+}| + |\psi_{+}\rangle \langle XX| + h.c.\big) . 
\label{h2rot}
\end{eqnarray}
Under the conditions $\label{eq:cond} |V_{F}|, |V_{XX}| \gg |\Omega^{'}|/2 $ the $|11\rangle \leftrightarrow |\psi_{+}\rangle$ and $|\psi_{+}\rangle \leftrightarrow |XX\rangle$ transitions are suppressed.

We use the quantum trajectories formalism to analyse the dynamics of our measurement process. As described in \cite{gardiner00,stace03} the conditional master equation (CME) 
for a system with $n$ imperfect measurement channels is:
\begin{eqnarray} \label{eq:CME}  d\rho_{c}  &=& -i[H,\rho_{c}] dt \nonumber \\ &&  + \sum_{j=1}^{n} \Big\{ (\eta_{j} Tr ( \mathcal{J}[c_j]\rho_c ) \rho_{c} + (1-\eta_j) \mathcal{J}[c_j] 
 \nonumber \\ && - \mathcal{A}[c_j]\rho_c ) dt + 
\Big(\frac{\mathcal{J}[c_j]\rho_c }{Tr(\mathcal{J}[c_j]\rho_c )} - \rho_{c}\Big) dN_{j} \Big\} ,  
\label{CME}
\end{eqnarray}
where $\rho_{c}$ is the density matrix of the system, $H$ is the system Hamiltonian in the interaction picture, $c_j$ is the Lindblad operator 
through which the system couples to the measurement channel $j$, $\mathcal{J}[c_j]$ is the jump operator which projects out the component of the state that 
is consistent with a detection in channel $j$ and is defined as $\mathcal{J}[c_j] \rho_c =c^{\dagger}_j \rho_c c_j$. $\mathcal{A}[c_j]$ is defined as 
$\mathcal{A}[c_j]\rho_c=\frac{1}{2} (c^{\dagger}_j c_j \rho_c + \rho_c c^{\dagger}_j c_j)$, $\eta_j$ is the efficiency of detector channel $j$ and $dN_j$ 
is the classical stochastic increment taking the values $\{0,1\}$ and denotes the number of photons detected from channel $j$ in the interval $t , t+dt$.
Eq.~\ref{CME} is equivalent to the linear, unnormalised, CME
\begin{equation} \dot{\tilde{\rho}} = - i [H,\tilde{\rho}] + \sum_{j}^{n} \big\{(1-\eta_{j}) \mathcal{J}[c_j]\tilde{\rho} - \mathcal{A}[c_j]\tilde{\rho}\big\} . \end{equation}   
where $\rho_c = \tilde{\rho}/Tr(\tilde{\rho})$.

For coupling strengths satisfying the criteria following Eq.~\ref{h2rot}, we are able to consider only one coupling channel describing the continuous 
measurement process. This coupling channel describes the radiative decay of the excited states in the odd-parity 
subspace. The coupling operator is taken to be of the form $c_{1} = \sqrt{\Gamma_{1}} (|10\rangle \langle X0| + |01\rangle \langle 0X|)$, where 
$\sqrt{\Gamma_{1}}$ is the decay rate for a single exciton, and the detector efficiency is $\eta_{1}$. This form of the Lindblad operator ensures that the measurement does not distinguish photons originating from different dots which, as we show later, is reasonable for sufficiently close QDs.

In order to characterize the time dependence of this relaxation, we consider typical interaction strengths $V_{F}=0.85~\text{meV}$ \cite{birkedal01} and 
$V_{XX}=5~\text{meV}$ \cite{biolatti02,lovett03b}, while a typical exciton creation energy is $\omega_{0}=2~\text{eV}$. We also
require that $\Omega \sim 0.1 ~\text{meV}$. The typical decay rate for an exciton in a QD has been measured to be $\tau_{X} \approx 1 ~\text{ns}$ 
\cite{borri01}, which gives a decay constant of $\Gamma_{1} \approx 4 ~\mu \text{eV}$.

We define the fidelity of the measurement, conditional on not measuring a photon, as $F_{0}=\langle \psi_{E} | \rho_f | \psi_{E} \rangle $ where $\rho_f$ is the state of the system at the end of the measurement
and the target state is $|\psi_{E}\rangle = \alpha_{00}|00\rangle + \alpha_{11} |11\rangle$. Meanwhile, the fidelity conditional on measuring the photon at time $t$ 
is $ F_{1}(t)=\langle \psi_{O} | \rho(t) | \psi_{O} \rangle$ with target state $|\psi_{O}\rangle = \alpha_{01}|01\rangle + \alpha_{10} |10\rangle$.

\begin{figure}[t]
\begin{center}
\includegraphics[width=.45\textwidth]{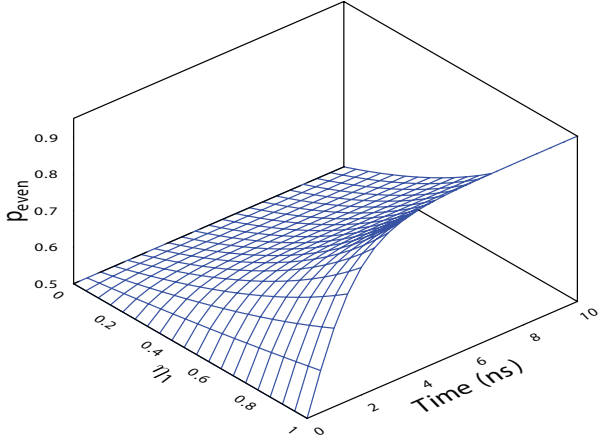}
\caption{\label{fig2} Probability that the state is in the even-parity subspace $p_{even}$ conditional on not measuring a photon.}
\end{center}
\end{figure}

We solve the CME analytically for our simple model, with the state initially in an equal superposition of all four computational basis states. The 
probability, $p_{even}$, that at a time $t$ we are in the even subspace conditioned on not observing a photon is:
\begin{equation} p_{even}(t) = \frac{1}{2+\eta_{1}(e^{-\Gamma_{1}t}-1)} , \end{equation}
while, on measuring a photon the probability that we have collapsed into the odd-space, $p_{odd}(t)$, is unity, as expected. The probability 
$p_{even}(t)$ is plotted in Fig. \ref{fig2} as a function of time and detector efficiency $\eta_{1}$.

After $10~\text{ns}$ the probability of photon emission is negligible, and there is essentially no further evolution 
beyond this time. This sets a characteristic time-scale for the measurement process. At this time if no photon has been measured we obtain a fidelity of $F_{0}=(2-\eta_{1})^{-1}$. Realistically, the losses at various stages in the detection procedure give rise to a photon detection efficiency of 
$50\%$ \cite{barnes02}; a value that will be used throughout the rest of the paper. To boost the fidelity we can repeat the excite-decay 
procedure several times.  Every additional cycle wherein no photon is detected increases our confidence that the state is in the even parity subspace. The fidelity after $n$ excite-decay cycles is $F_{0}^{n} = [1+(1-\eta)^{n}]^{-1}$.

We have shown above that on detection of a photon, we obtain a perfect fidelity for our parity measurement. However, we have up to this point neglected a number 
of effects that may reduce the performance of our measurement. For example, we have ignored the effects of excitations to the $|\psi_{+}\rangle$ and 
$|XX\rangle$ states. We have also ignored two possible ways in which the QDs could be distinguished when a photon is detected, namely spatial separation and 
non-resonance of the QDs. The effects of these three factors will be analysed in the following sections.

To analyze the effect of spatial separation we derive the CME from first principles, starting from the microscopic Hamiltonian: $H_{tot} = \omega_{0}(c_{A}^{\dagger} c_{A} + c_{B}^{\dagger} c_{B}) + \sum_{k} \omega_{k} a^{\dagger}_{k} a_{k} + H_{I}$, where
\begin{equation} H_{I} =  \sum_{\mathbf{k}} (\mathbf{\mu}.\mathbf{\hat{\sigma}_{\mathbf{k}}}) \epsilon_{\mathbf{k}} a_{\mathbf{k}} e^{i \mathbf{k}.\mathbf{r}} (c^{\dagger}_{A} + e^{i  \mathbf{k}.\Delta\mathbf{r}} c^{\dagger}_{B}) + h.c.  \end{equation}
$\mu$ is the dipole moment vector for each qubit, $\hat{\sigma}_{\mathbf{k}}$ is the polarisation vector for the electric field, $a_{\mathbf{k}}$ is the 
annihilation operator for a quantum of the electric field, $\mathbf{k}$ is the wavevector for the electric field, $c_{A,B}$ represents the 
annihilation operator for an exciton on dot $A,B$ respectively and $\Delta\mathbf{r}$ is the center-to-center separation of the dots.

To proceed we iterate the Schroedinger Equation twice to 
obtain an integro-differential equation describing the evolution. Then we trace out the environmental degrees of freedom to obtain an unconditional master 
equation. Finally we generate a CME by defining a jump operator describing the detection process. This yields the following:
\begin{equation} \dot{\tilde{\rho}} = -i[H,\tilde{\rho}] + \sum_{\mathbf{k}} (1-\eta_{\mathbf{k}}) P_{\mathbf{k}} \tilde{\rho} P_{\mathbf{k}}^{\dagger} - \{P_{\mathbf{k}}^{\dagger} P_{\mathbf{k}} \tilde{\rho} + \tilde{\rho} P_{\mathbf{k}}^{\dagger} P_{\mathbf{k}}\} , \end{equation}
where $P_{\mathbf{k}}^{\dagger}(r) = (\mathbf{\mu}.\mathbf{\hat{\sigma}_{\mathbf{k}}}) \epsilon_{\mathbf{k}} e^{i \mathbf{k}.\mathbf{r}}(c_{A}^{\dagger} +e^{i \mathbf{k}.\Delta\mathbf{r}} c_{B}^{\dagger})$.
Summing over all the modes, we obtain:
\begin{equation} \dot{\tilde{\rho}} = -i[H,\tilde{\rho}] + (1-\eta ) \mathcal{J}\tilde{\rho} - \mathcal{A}\tilde{\rho} \end{equation} 
where 
\begin{eqnarray} &&\mathcal{J}\tilde{\rho} = \Gamma_{1} [c_{A} \rho c_{A}^{\dagger} + 3 f(k_{0}\Delta r) (c_{A} \rho c_{B}^{\dagger} + c_{B} \rho c_{A}^{\dagger}) + c_{B} \rho c_{B}^{\dagger}],  
\nonumber\\ &&\mathcal{A}\tilde{\rho} = \Gamma_{1} [c_{A}^{\dagger} c_{A} \rho + \rho c_{A}^{\dagger} c_{A} + c_{B}^{\dagger} c_{B} \rho + \rho c_{B}^{\dagger} c_{B}], 
\nonumber\\ &&f(\alpha) = \frac{2\alpha\cos(\alpha) + (\alpha^2 -2)\sin(\alpha)}{\alpha^3} .
\end{eqnarray}
Hence the fidelity when a photon is detected is
\begin{equation} F = \frac{1 + 3 f(k_{0}\Delta r)}{2}. \end{equation}
In order for our scheme to work the Foerster interaction strength must be of order 1~meV~\cite{lovett03b}, and this sets a value for $\Delta r$ of  $5~\text{nm}$~\cite{lovett03b}. Using $\omega_{0}=2~\text{eV}$ we obtain $k_{0} = 10^{7}~m^{-1}$, and $k_{0} \Delta r = 5 \times 10^{-2}$. This leads to a modified fidelity of $0.999$ -- and thus we conclude that our scheme is resilient to 
effects of spatial separation.

We now consider the situation of non-resonant QDs. If the detuning of the two transitions $\delta=\omega_a - \omega_b$ is large enough it will destroy the delocalization 
and resulting splitting due to the Foerster interaction, thus preventing selective excitation to states only in the odd subspace. Two inequalities must be satisfied: first, in order to excite excitons from both $\ket{10}$ and $\ket{01}$ with a single laser pulse, we require that $\delta \ll \Omega$. Second, to restrict transitions in the $\mathcal{H}_{2}$ subspace, we require that $\sqrt{\delta^{2} + V_{F}^{2}}, V_{XX} \gg \Omega (b_{2} \pm b_{1})$
where $b_{1,2} = \sqrt{\frac{A \mp 1}{2A}}$ and $A=\sqrt{1+\frac{V_{F}^{2}}{\delta^{2}}}$. These lead to the condition that $V_F \gg \delta$.

Returning to the effects of non-resonant dots on the relaxation process, we calculate a modified CME using the same method as used for the case of spatially separated dots. 
We obtain $c_{1} = \sqrt{\Gamma_{1}} (c_{A} + e^{-i \delta t} c_{B})$.  Using this modified Lindblad operator in the CME we find that on measuring a photon the state of the system becomes:
\begin{equation} |\psi(t)\rangle =  \alpha_{01}|01\rangle + \alpha_{10} e^{i\delta t}|10\rangle . \end{equation}
This extra phase is in general unknown and so is detrimental to the parity measurement, since it destroys the coherence between the states. However, accurate timing of the photon detection corrects for this; we can reverse the (now known) accumulated phase using single qubit phase gates that can be implemented optically~\cite{lovett06}. 
State-of-the-art photon detectors have time resolutions of the order of picoseconds \cite{picoquant}, so it is possible to correct for detunings of the order of 1~meV.  This regime can be achieved with existing technology using an electric field to Stark shift the QDs on to 
resonance~\cite{nazir05}. Alternatively, we could use molecular systems \cite{hettich02} 
which are identical and so the problem of non-resonance is 
effectively eliminated.

\begin{figure}[t]
\begin{center}
\includegraphics[width=.4\textwidth]{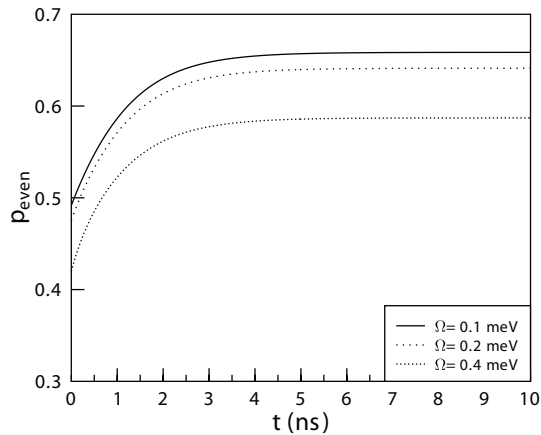}
\caption{\label{fig3} $p_{even}$ for various laser coupling strengths, $\Omega$, when photons from $\mathcal{H}_{2}$ subspace are not detectable but simply lost to the environment.}
\end{center}
\end{figure}

Finally we consider the potential problem of excitations in the $\mathcal{H}_{2}$ subspace. To model these we allow a further two decay channels: $c_{2}=\sqrt{\Gamma_{2}}|11\rangle\langle\psi_{+}|$ 
and $c_{3}=\sqrt{\Gamma_{3}}|\psi_{+}\rangle\langle XX|$. The decay rates for these two channels are set by the dipole moments for the transitions. The allowed transitions within the $\mathcal{H}_2$ subspace have a larger dipole hence
$\sqrt{\Gamma_{2}} = \sqrt{\Gamma_{3}} = \sqrt{2 \Gamma_{1}}$. We will assume that photons from these extra channels can be filtered out before they reach
the detector. Numerical simulations for the probability that the system is in the even-parity 
subspace are presented in Fig. \ref{fig3}. Although there is some degradation in the performance, even for the strongest displayed laser coupling we obtain a final 
fidelity of over $0.5$. This is sufficient to enable us to obtain extremely high fidelity using only a few rounds of the excite-decay procedure. 

Spin relaxation effects (including electron-hole interactions) couple the subspaces and so might be a problem for our protocol. However, they occur on a timescale of the order of 20 ms~\cite{kroutvar04}, which can be ignored on the timescale of our measurement. Further, the electron spin-spin exchange interaction has been measured at less than $1~\mu \text{eV}$ in a quantum dot system under a range of conditions~\cite{laird06}. This is considerably weaker than any other interactions present and can thus be safely neglected. 

We have described a reliable method of performing a spin-parity measurement via the detection of a photon. Beenakker 
\textit{et al.} suggested that this could be used to construct a CNOT gate by arranging two parity measurement gates in parallel. This is possible in a chain 
of three QDs if we have the ability to address two of the QDs with the laser while leaving the final QD unaffected. This may be achieved by using two 
different exciton transitions for the two different entangling gates. 

The entangling procedure that we have described could be incorporated into a scheme to 
grow large scale graph states reliably \cite{benjamin05a}. Graph states are a certain type of multi-entangled state, which enable one to perform computational 
operations purely by performing single-qubit measurements. A major difficulty with graph state computation comes from the successful preparation of the
initial multi-entangled state. Benjamin {\it et al.}~\cite{benjamin05a} propose a method of overcoming this in systems like the one discussed here, where a reliable method of entangling between pairs of qubits exists. Different pairs are then linked through any entangling process (that may be inefficient).

In conclusion, we have presented a novel scheme for implementing a spin-parity measurement on a pair of coupled quantum dots. We have estimated the fidelity of the parity measurement scheme presented here, and found it to be robust ($F>95\%$) in the presence of realistic sources of errors, including inefficient photon-detection, unwanted excitations in the $\mathcal{H}_2$ subspace, and spatial or spectral separation of the QDs.
Finally, we identified two applications for our parity measurement: an implementation of a CNOT gate and a method of reliably constructing large scale 
graph states.

\begin{acknowledgements}
This work is supported by the QIPIRC www.qipirc.org (GR/S82176/01). BWL is supported by DSTL and St Anne's College, Oxford. BWL and SCB acknowledge support from the Royal Society.
\end{acknowledgements}

\end{document}